\documentclass[aps,twocolumn,nofootinbib,notitlepage,longbibliography,superscriptaddress]{revtex4-2}

\usepackage{amsmath,amssymb,amsfonts}
\usepackage{physics}
\usepackage{bm}
\usepackage{float}
\usepackage[T1]{fontenc}
\usepackage[utf8]{inputenc}
\usepackage{graphicx}
\usepackage{booktabs}
\usepackage[colorlinks,linkcolor=blue,citecolor=blue,urlcolor=blue]{hyperref}
\usepackage{subcaption}
\usepackage{ragged2e}
\usepackage[compatibility=false]{caption}
\usepackage{fancyhdr}
\usepackage{xcolor}

\DeclareCaptionJustification{justified}{\justifying}
\definecolor{notetext}{rgb}{0.7,0,0}
\definecolor{revtext}{rgb}{0,0,1}
\newif\ifnotes \notestrue
\newif\ifrev   \revtrue

\DeclareMathOperator{\arcsinh}{arcsinh}
\DeclareMathOperator{\arctanh}{arctanh}
\renewcommand{\Re}{\mathrm{Re}\,}
\renewcommand{\Im}{\mathrm{Im}\,}

\renewcommand{\d}{\mathrm{d}}
\newcommand{\half}{\tfrac{1}{2}}

\newcommand{\phivac}{\phi_{\mathrm{vac}}}
\newcommand{\mbare}{m_{\mathrm{bare}}}
\newcommand{\mren}{m_{\mathrm{ren}}}
\newcommand{\Mcl}{M_{\mathrm{cl}}}
\newcommand{\Msol}{M_{\mathrm{sol}}}
\newcommand{\Mqp}{M_{\mathrm{QP}}}
\newcommand{\Mzam}{M_{\mathrm{Zam}}}
\newcommand{\Csch}{\mathcal{C}}

\newcommand{\normord}[1]{\mathopen{:}\!#1\!\mathclose{:}}

\newcommand{\dxphys}{\Delta x_{\mathrm{phys}}}
\newcommand{\dxcm}{\Delta x_{\mathrm{cm}}}
\newcommand{\Qpos}{Q_+}

\newcommand{\secref}[1]{Sec.~\ref{#1}}
\newcommand{\figref}[1]{Fig.~\ref{#1}}
\newcommand{\tabref}[1]{Table~\ref{#1}}
\newcommand{\appref}[1]{Appendix~\ref{#1}}

\begin{document}

\title{Sine-Gordon Model with Bosonic Tensor Networks:\\ Continuum Matching and Soliton Scattering}
\author{Florian Hechenberger}
\email{florian.hechenberger@stonybrook.edu}
\author{Tommaso Rainaldi}
\email{tommaso.rainaldi@stonybrook.edu}
\author{Felix Ringer}
\email{felix.ringer@stonybrook.edu}
\affiliation{Center for Nuclear Theory, Department of Physics and Astronomy, Stony Brook University, Stony Brook, New York 11794-3800, USA}

\date{\today}
\begin{abstract}
We use bosonic tensor networks to connect the lattice sine-Gordon model in the Hamiltonian formulation quantitatively to its continuum theory. Matching the lattice vertex operator to its conformal normalization at the free-boson ultraviolet fixed point yields the exact relation between the bare lattice coupling and the renormalized continuum mass parameter. The soliton mass then approaches Zamolodchikov's exact continuum prediction throughout the studied parameter range, without adjustable parameters. Using uniform matrix product states and a quasiparticle ansatz, we also recover the relativistic soliton dispersion and the two lightest breather masses at the percent level. We simulate real-time collisions of Gaussian soliton-antisoliton wave packets near a reflectionless point and extract the Wigner spatial displacement. We compare this displacement with the exact continuum prediction obtained from the momentum derivative of the transmission phase, recovering its characteristic rapidity dependence. Our bosonic simulations provide a foundation for nonintegrable extensions, offer lessons for renormalization in other Hamiltonian lattice field theories, including gauge theories, and provide benchmarks for continuous-variable quantum simulations.
\end{abstract}
\maketitle

\tableofcontents

\section{Introduction}
\label{sec:intro}

The sine-Gordon model in $1+1$ dimensions provides a setting in
which the masses and interactions of topological particles can
be studied nonperturbatively and compared with exact analytical
predictions. Its periodic potential supports solitons and
antisolitons, which interpolate between neighboring vacua,
together with neutral bound states known as breathers in the
attractive regime. These excitations connect the structure of
the vacuum to observable particle properties and scattering
dynamics. Integrability determines the factorized scattering
amplitudes~\cite{Zamolodchikov:1978xm}, while an exact
mass-coupling relation connects the soliton mass to the
parameters of the continuum theory~\cite{Zamolodchikov:1995xk}.
The model therefore offers complementary tests of numerical
approaches through its spectrum and real-time dynamics.

The relevance of sine-Gordon theory extends beyond this role
as an exactly solvable benchmark. Through bosonization, it is
equivalent to the massive Thirring model~\cite{Coleman:1974bu},
and it describes effective baryon dynamics in large-$N$
two-dimensional
QCD~\cite{Witten:1979kh,Steinhardt:1980ry,Florio:2022uvd,Cogburn:2026aqy}.
Moreover, exact knowledge of the spectrum and scattering
amplitudes does not make general correlation functions or
the evolution of spatially inhomogeneous states readily
accessible. These remain active subjects of research,
including at finite temperature and out of
equilibrium~\cite{Horvath:2021vlx,Kormos:2022rtu,Toth:2026yyj}.

Recovering continuum predictions from a spatially discretized
Hamiltonian requires control over several distinct
approximations. The bare parameters must be related to a
specified continuum normalization, while the local bosonic
Hilbert space and the entanglement retained in the numerical
state must both be truncated. Reducing the lattice spacing
therefore needs to be accompanied by control of these
truncations. Establishing this connection between the
regulated Hamiltonian and its physical observables is
particularly valuable in an integrable theory, where exact
results can validate methods intended for problems without
analytical solutions.

We address these questions using bosonic matrix product
states (MPS)~\cite{Vanderstraeten:2019voi}, representing the
scalar field and its conjugate momentum in a local oscillator
basis. This representation retains the bosonic field
variables explicitly and allows the local Fock cutoff to be increased systematically. A related approach
has recently been applied to false vacuum decay in scalar
field theory~\cite{Abel:2025pxa}. Uniform MPS describe the
interacting vacua directly in the thermodynamic
limit~\cite{Zauner-Stauber:2017eqw}, while a quasiparticle
ansatz gives access to solitons, antisolitons, and
breathers~\cite{Vanderstraeten:2019voi,Milsted:2020jmf}.
Localized excitations can then be evolved in real time
using the time-dependent variational
principle (TDVP)~\cite{Haegeman:2011zz,Haegeman:2014gua}.
The same bosonic Hamiltonian thus provides access to both
the particle spectrum and scattering dynamics.

Real-time wave packet scattering with MPS has been studied
in Ising field theory~\cite{Jha:2024jan}, scalar
$\phi^4$ theory~\cite{Sayegh:2025rlp}, and two-dimensional
quantum
electrodynamics~\cite{Rigobello:2021fxw,Belyansky:2023rgh}.
Related methods have also been developed for collisions
of quasiparticles and topological domain-wall
excitations~\cite{VanDamme_2021,Milsted:2020jmf}.
Wave packet scattering has also been brought to quantum
hardware and analog simulators: time-delay extraction on
trapped ions and superconducting qubits~\cite{Gustafson:2021imb},
hadron wave packet preparation and propagation in the
Schwinger model~\cite{Farrell:2024fit,Davoudi:2024wyv},
fermion wave packet scattering in the Thirring
model~\cite{Chai:2023qpq}, the fermionic dual of the model
studied here, and meson collisions on spin and cold-atom
platforms~\cite{Surace:2020ycc,Bennewitz:2025nhz,Su:2024uuc}. 
Our study brings together
an explicit bosonic representation of the sine-Gordon
Hamiltonian, analytically determined lattice-to-continuum
matching, and exact continuum predictions for both the
spectrum and soliton scattering.

The continuum matching is obtained by comparing the
short-distance normalization of lattice and continuum
vertex operators at the free-boson ultraviolet fixed point.
This fixes the matching constant exactly for the lattice
regularization and continuum normalization adopted here,
independently of the numerically measured spectrum.
Combined with Zamolodchikov's mass-coupling
relation~\cite{Zamolodchikov:1995xk}, it yields a prediction
for the continuum soliton mass without adjustable parameters.
The numerical soliton mass approaches this prediction
throughout the studied parameter range. We also obtain
the two lightest breather masses and examine the recovery
of the relativistic soliton dispersion as the lattice
spacing is reduced. These comparisons test the matching
and the numerical representation across several
independent observables.

Real-time scattering probes information beyond the particle
spectrum. We prepare spatially separated Gaussian soliton
and antisoliton wave packets and study their collisions
near a reflectionless point of the continuum theory.
At the reflectionless point, the exact scattering amplitude
predicts pure transmission, but the interaction still
displaces the outgoing wave packets relative to freely
propagating reference states. This Wigner spatial
displacement is determined by the momentum derivative
of the transmission phase~\cite{Wigner:1955zz}, providing
a direct connection between the simulated trajectories
and the continuum scattering amplitude.

We extract the displacement by tracking the topological
charge density and compare it with the exact continuum
prediction averaged over the momentum distribution of
the wave packets. The simulations reproduce the expected
transmission behavior and the rapidity dependence of the
displacement. The remaining quantitative deviations
motivate checks of the sensitivity to wave packet
preparation, finite evolution time, and numerical
truncations.

Our bosonic simulations provide a starting point for
studying nonintegrable deformations, where exact scattering
amplitudes are generally unavailable. The explicit
separation of continuum matching from the calculation
of physical observables also offers methodological guidance
for renormalization in other Hamiltonian lattice field
theories, including gauge theories. The necessary matching
conditions and counterterms must, however, be established
for each theory and its regularization.

Quantum simulation of quantum field theories has developed into a broad
program~\cite{Jordan:2012xnu,PRXQuantum.4.027001,Bauer:2023qgm,PRXQuantum.5.037001}, with scalar
field theory as the standard testbed for bosonic degrees of freedom on
qubits~\cite{Jordan:2012xnu,Klco:2018zqz,Yeter-Aydeniz:2018mix}.
The local oscillator representation also connects directly
to continuous-variable quantum simulation, where bosonic
modes encode the field and its conjugate
momentum~\cite{Marshall:2015mna,Briceno:2023xcm,Stavenger:2022wzz,Liu:2024mbr,Araz:2024dcy,Kemper:2025ldr}.
Trigonometric continuous-variable gates provide a way to
implement cosine
interactions~\cite{Rainaldi:2025ymn,Chalermpusitarak:2025cod}.
Recent work on hybrid qubit-qumode architectures has
studied vacuum preparation, time-dependent correlation
functions, and soliton profiles for the same lattice
Hamiltonian~\cite{Rainaldi:2025ymn}. Our calculations extend
these studies to larger systems, soliton scattering, and
quantitative continuum comparisons, providing classical
benchmarks for the preparation and evolution of
interacting bosonic states.

The remainder of this work is organized as follows.
\secref{sec:model} introduces the continuum theory and its
exact spectrum and scattering amplitudes.
\secref{sec:lattice} presents the lattice Hamiltonian,
the local oscillator basis, and the matching of the lattice
simulation to continuum results. \secref{sec:results} reports the numerical
results for the spectrum, dispersion, and soliton-antisoliton
scattering. \secref{sec:conclusions} summarizes the findings
and discusses future directions. The tensor network
constructions are collected in \appref{app:tn}.

\section{The Sine-Gordon Model}
\label{sec:model}

In this section, we define the continuum model, the conventions used throughout this work, and collect the
exact continuum results that the lattice simulations in \secref{sec:results} will be compared against. We work with the $(1+1)$-dimensional sine-Gordon Lagrangian density
\begin{equation}
  \mathcal{L}
  = \half(\partial_\mu\phi)^2
  + \frac{m^2}{\beta^2}\bigl(\cos(\beta\phi) - 1\bigr),
  \label{eq:lagrangian}
\end{equation}
where $\phi(x,t)$ is a real scalar field, $m$ is the bare mass parameter of mass dimension one,
and $\beta$ is a dimensionless coupling. The potential $V(\phi) = (m^2/\beta^2)(1-\cos(\beta\phi))$
has period $2\pi/\beta$ and vanishes at its minima $\phivac^{(n)} = 2\pi n/\beta$,
$n\in\mathbb{Z}$. The semiclassical regime, in which the classical soliton picture is quantitatively
reliable, corresponds to $\beta^2 \ll 8\pi$. The quantum theory undergoes a
Berezinskii-Kosterlitz-Thouless (BKT) transition at
$\beta^2 = 8\pi$~\cite{Berezinskii:1970pzv,Berezinskii:1972fet,Kosterlitz:1973xp,Coleman:1974bu},
above which the cosine is irrelevant, and the model flows to a free massless boson.

The vacuum manifold $\{\phivac^{(n)}\}_{n\in\mathbb{Z}}$ is a discrete set related by the shift
symmetry $\phi\to\phi + 2\pi/\beta$. Field configurations that interpolate between different
vacua as $x\to\pm\infty$ carry a conserved topological charge
\begin{equation}
  Q = \frac{\beta}{2\pi}\int_{-\infty}^{\infty}\d x\,\partial_x\phi
    = \frac{\beta}{2\pi}\bigl[\phi(+\infty)-\phi(-\infty)\bigr]
    \in \mathbb{Z}.
  \label{eq:topcharge}
\end{equation}
Configurations with $Q=\pm1$ are called solitons and antisolitons, respectively. This charge is the
observable used to track the two solitons through the collision in
\secref{sec:results:scattering}, since it distinguishes them at every time step, even when the energy density does not.

Several different normalizations appear in the literature. \tabref{tab:conventions} translates
between the convention used here and the one adopted by Ref.~\cite{Lukyanov:1996jj}, whose
mass formula we compare against.

\begin{table*}[t]
  \centering
  \begingroup
  \small
  \setlength{\tabcolsep}{6pt}
  \begin{tabular*}{\textwidth}{@{\extracolsep{\fill}}lcc@{}}
    \toprule
    Quantity & This work & Ref.~\cite{Lukyanov:1996jj} \\
    \midrule
    Field
      & $\phi$
      & $\phi_Z=\sqrt{8\pi}\,\phi$ \\
    Coupling
      & $\beta$
      & $\beta_Z=\beta/\sqrt{8\pi}$ \\
    Mass parameter
      & $m^2$
      & $\mu=m^2/(2\beta^2)$ \\
    Lagrangian
      & $\tfrac{1}{2}(\partial_\mu\phi)^2
         +\tfrac{m^2}{\beta^2}\bigl(\cos(\beta\phi)-1\bigr)$
      & $\tfrac{1}{16\pi}(\partial_\mu\phi_Z)^2
         -2\mu\cos(\beta_Z\phi_Z)$ \\
    Semiclassical limit
      & $\beta^2\to0$
      & $\beta_Z^2\to0$ \\
    BKT transition
      & $\beta^2=8\pi$
      & $\beta_Z=1$ \\
    \bottomrule
  \end{tabular*}
  \endgroup
  \caption{Conversion between our conventions and those of
    Ref.~\cite{Lukyanov:1996jj}, which uses a Euclidean convention.}
  \label{tab:conventions}
\end{table*}

The classical soliton solution centered at $x_0$ is
\begin{equation}
  \phi_S(x; x_0) = \frac{4}{\beta}\arctan\!\bigl(e^{m(x-x_0)}\bigr),
  \label{eq:solitonprofile}
\end{equation}
interpolating from $\phi_S(-\infty)=0$ to $\phi_S(+\infty)=2\pi/\beta$, which is $Q=+1$ in
Eq.~\eqref{eq:topcharge}. The antisoliton profile of charge $Q=-1$ follows by a shift
\begin{equation}
  \phi_{\bar S}(x; x_0)
  = \frac{2\pi}{\beta} - \phi_S(x; x_0).
  \label{eq:antisolitonprofile}
\end{equation}
The classical rest-mass energy is
\begin{equation}
  \Mcl = \frac{8m}{\beta^2}.
  \label{eq:classicalmass}
\end{equation}
Since the soliton width is set by $m^{-1}$, resolving the profile on a lattice requires
$am \ll 1$. The lattice spacings used in \secref{sec:results} are chosen such that this condition is well satisfied. 
The exact quantum soliton mass is known~\cite{Zamolodchikov:1978xm} and reads, in our
conventions,
\begin{align}
  &\Msol
  = \nonumber\\
  &\frac{2}{\sqrt{\pi}}
    \frac{\Gamma\!\bigl(\xi/2\bigr)}{\Gamma\!\bigl((1+\xi)/2\bigr)}
    \left(
      \frac{\pi\mren^2}{2\beta^2}
      \frac{\Gamma\!\bigl(1 - \beta^2/(8\pi)\bigr)}
           {\Gamma\!\bigl(\beta^2/(8\pi)\bigr)}
    \right)^{\frac{1}{2 - \beta^2/(4\pi)}}.
  \label{eq:exactmass}
\end{align}
Here,
\begin{equation}
  \xi = \frac{\beta^2}{8\pi - \beta^2}\,,
  \label{eq:xi}
\end{equation}
and $\mren$ is the renormalized mass obtained after normal-ordering the cosine potential, defined
in \secref{sec:renorm}. Note that Eq.~\eqref{eq:exactmass} is written in terms of the renormalized mass $\mren$, rather than the bare parameter $\mbare$. In the limit $\beta^2\to 0$, the classical
result~\eqref{eq:classicalmass} is recovered. Both the normal-ordering factor and the mass
formula hold to all orders in the coupling, such that their combination gives the physical
continuum mass without perturbative approximations.

The breather modes, neutral soliton-antisoliton bound states, have masses~\cite{Dashen:1975hd}
\begin{equation}
  M_n = 2\Msol\sin\!\left(\frac{n\pi\xi}{2}\right),
  \qquad n = 1, 2, \ldots, \left\lfloor \frac{1}{\xi} \right\rfloor,
  \label{eq:breathermasses}
\end{equation}
such that the number of bound states is $\lfloor 1/\xi\rfloor$ and every breather satisfies
$M_n \leq 2\Msol$ identically. For $\xi\geq 1$, equivalently $\beta^2 \geq 4\pi$, no breather
exists, and the interaction between soliton and antisoliton is repulsive. 
Conventionally, the lightest breather is usually referred to as the meson.

Scattering in the soliton sector is elastic, and the two-body $S$-matrix is block
diagonal~\cite{Zamolodchikov:1978xm}, see also Ref.~\cite{Mussardo:2020rxh}\footnote{The
definition of $\xi$ in Ref.~\cite{Mussardo:2020rxh} differs from ours by a factor of $\pi$.},
\begin{equation}
    \mathcal{S}(\theta)=
    \begin{pmatrix}
        S & 0 & 0 & 0\\
        0 & S_T & S_R & 0 \\
        0 & S_R & S_T & 0 \\
        0 & 0 & 0 & S
    \end{pmatrix}.
\end{equation}
Here,
\begin{equation}
\label{eq:zam_s_mat}
    S(\theta)=-\exp\left[-i
    \int_0^\infty\frac{\mathrm{d}t}{t}\frac{\sinh\frac{\pi(1-\xi)t}{2}}{\sinh\frac{\pi\xi t}{2}\cosh{\frac{\pi t }{2}}}\sin\theta t
    \right],
\end{equation}
and $\theta$ is the relative rapidity of the two-particle system,
$\theta = \arctanh\!\left[(v_S-v_{\bar S})/(1-v_S v_{\bar S})\right] = 2\theta_S = -2\theta_{\bar S}$.
Factorization, in the form of the Yang-Baxter equation satisfied by the amplitudes, fixes
the transmission and reflection components relative to the soliton-soliton amplitude
$S(\theta)$, with
\begin{equation}
    \label{eq:phase_tr}
    S_T(\theta)=\frac{\sinh\frac{\theta}{\xi}}{\sinh{\frac{i\pi-\theta}{\xi}}}S(\theta),\quad  S_R(\theta)=i\frac{\sin\frac{\pi}{\xi}}{\sinh{\frac{i\pi-\theta}{\xi}}}S(\theta).
\end{equation}
The overall factor $S(\theta)$ of Eq.~\eqref{eq:zam_s_mat} is not fixed by factorization
alone. Unitarity and crossing symmetry constrain it only up to a Castillejo-Dalitz-Dyson
(CDD) factor~\cite{Castillejo:1955ed}. Each CDD factor adds a pole to the physical strip
$0 < \Im\,\theta < \pi$, where a pole signals a bound state, and
Eq.~\eqref{eq:zam_s_mat} is the minimal solution, carrying no poles there beyond those required
by the breather spectrum of Eq.~\eqref{eq:breathermasses}.
At $\xi = 1/3$, that is $\beta^2 = 2\pi$, the reflection amplitude vanishes and the scattering
is purely transmissive, although introducing a global phase shift to the scattered states. The same happens at every $\xi = 1/n$, where
Eq.~\eqref{eq:zam_s_mat} can be evaluated in closed form
\begin{equation}
\label{eq:zam_s_mat_analytic}
    S_T(\theta)\big|_{\xi=1/n}=e^{in\pi}\prod_{k=1}^{n-1}\frac{e^{\theta-ik\pi/n}+1}{e^\theta + e^{-ik\pi/n}}.
\end{equation}
The point $\xi = 1/3$ is also distinguished spectroscopically, since the first breather mass in Eq.~\eqref{eq:breathermasses} then coincides with the soliton mass. We therefore perform the scattering simulations of \secref{sec:results:scattering} at this point, where the presence of a single outgoing channel avoids the need to disentangle transmitted and reflected contributions in the measured charge density.

MPS methods are blind to global phases, which is why we cannot directly test our numerical results against theory predictions for it. On the other hand, the elastic phase shift additionally manifests itself as a spatial displacement of the outgoing trajectories, relative to free propagation. For the transmitted soliton, the Wigner relation~\cite{Wigner:1955zz} gives the following result for the asymptotic displacement of the outgoing soliton or antisoliton
\begin{equation}
\label{eq:displacement_analytical}
    \Delta x_{S/\bar{S}}
    = -\frac{\partial\arg S_T(\theta_{12})}{\partial p_{S/\bar{S}}}
    = \mp\frac{1}{M\cosh\theta_{S/\bar{S}}}
       \frac{\d\arg S_T}{\d\theta_{12}},
\end{equation}
where $\theta_{12} = \theta_S - \theta_{\bar{S}}$ is the relative rapidity of the pair and
$E_{S/\bar{S}} = M\cosh\theta_{S/\bar{S}}$ is the single-particle energy. The opposite signs for soliton and antisoliton follow from
$\partial\theta_{12}/\partial p_S = 1/E_S$ and $\partial\theta_{12}/\partial p_{\bar S} = -1/E_{\bar S}$.
Equation~\eqref{eq:displacement_analytical} applies to momentum eigenstates, whereas the states used in the simulations are wave packets of finite width. The continuum prediction must therefore be averaged over the momentum distribution of the wave packet. We construct this averaged prediction in \secref{sec:results:scattering}, where we also specify the wave packet profile.

\section{The Lattice Sine-Gordon Model}
\label{sec:lattice}

Here we formulate the lattice Hamiltonian in the local basis in \secref{sec:ham} and describe the fit used to extract the soliton mass from its dispersion relation. This form of the Hamiltonian is the starting point for the simulations described in \secref{sec:results}. In \secref{sec:renorm}, we then relate the lattice Hamiltonian to the continuum theory and describe the numerical matching procedure employed in this work.

\subsection{Hamiltonian}
\label{sec:ham}

We discretize the continuum theory on a one-dimensional spatial lattice with sites $j = 1,\ldots,N$ and
lattice spacing $a$~\cite{Kogut:1974ag}, which leads to
\begin{align}
  &H = \nonumber\\
  &\sum_{j=1}^{N}
  \left[
    \frac{1}{2a}\Pi_j^2
    + \frac{1}{a}\phi_j^2
    - \frac{1}{a}\phi_j\phi_{j+1}
    + \frac{a\mbare^2}{\beta^2}\bigl(1 - \cos(\beta\phi_j)\bigr)
  \right].
  \label{eq:latticeH}
\end{align}
Here, $\phi_j$ and $\Pi_j$ are canonically conjugate fields satisfying $[\phi_j,\Pi_k]=i\delta_{jk}$. Expanding the cosine to quadratic order gives the free lattice Hamiltonian
\begin{equation}
  H_0 = \sum_j\left[
    \frac{1}{2a}\Pi_j^2 + \frac{1}{a}\phi_j^2 - \frac{1}{a}\phi_j\phi_{j+1} + \frac{1}{2}a \mbare^2 \phi_j^2
  \right],
\end{equation}
which is a chain of coupled harmonic oscillators with dispersion relation
\begin{equation}
  \omega_k^2(a) = \frac{4}{a^2}\sin^2\!\!\left(\frac{ka}{2}\right) + \mbare^2.
  \label{eq:freedispersion}
\end{equation}
In the continuum limit, $a$ tends to zero at fixed $k$ and we obtain the relativistic dispersion relation $\omega^2\to k^2 + \mbare^2$. The vacuum sectors of
Eq.~\eqref{eq:latticeH} are translationally invariant and can therefore be represented directly in the thermodynamic limit using infinite MPS. The vacuum sector is fixed by the choice of asymptotic field value, while configurations interpolating between different vacua belong to topological sectors characterized by their conserved charge.

At each lattice site, the local Hilbert space is represented in the bosonic Fock basis $\{\ket{n}\}_{n=0}^{d-1}$,
with local dimension $d$~\cite{Abel:2025pxa,Sayegh:2025rlp}. The truncation of bosonic modes to a
finite Fock space and its convergence have been studied in the context of qubit
encodings~\cite{Macridin:2018gdw,Macridin:2018oli,Macridin:2021uwn,Klco:2018zqz}. The field and conjugate momentum operators are expressed through ladder operators as
\begin{equation}
  \phi_j = \frac{a_j + a_j^\dagger}{\sqrt{2}},
  \qquad
  \Pi_j =i \frac{(a_j^\dagger - a_j)}{\sqrt{2}}.
  \label{eq:ladderops}
\end{equation}
The cosine operator is obtained by matrix exponentiation in this truncated basis, $\cos(\beta\phi_j) = \Re\bigl[e^{i\beta\phi_j}\bigr]$, which is exact up to the truncation at
$d$ levels. Every result below must therefore be checked for convergence with respect to four quantities:
the lattice spacing $a$, the local Fock cutoff $d$, the MPS bond dimension $\chi$, and, once wave packets
are formed, the chain length $N$.

For a soliton with Bloch momentum $p\in[-\pi/a,\pi/a]$, we fit the numerically computed dispersion relation to the lattice
form
\begin{equation}
  E^2(p) = M^2 + C\sin^2\!\!\left(\frac{pa}{2}\right),
  \label{eq:solitondispersion}
\end{equation}
from which we extract the soliton rest mass $M$ and the velocity coefficient $C$. In the free-field limit
$C = 4/a^2$, such that $c_{\mathrm{lat}}^2 = Ca^2/4 \to 1$ in the continuum limit. A deviation of
$c_{\mathrm{lat}}$ from unity at finite $a$ is therefore a measure of lattice discretization effects.

\subsection{Renormalization}
\label{sec:renorm}

The cosine is a composite operator with scaling dimension $2\alpha$ ($\alpha=\beta^2/8\pi$) at the
free-boson ultraviolet fixed point, and its normalization on the lattice depends on the UV cutoff $a$. A comparison with the exact mass formula therefore requires
matching the lattice cosine to the conformal normalization in which Eq.~\eqref{eq:exactmass} is
written. In this section, we fix the normalization exactly, which allows for a parameter-free comparison to the results in the continuum, see \secref{sec:results:spectrum}. The normal-ordered cosine is related
to the corresponding bare quantity as
\begin{equation}
  \cos(\beta\phi) = Z_1^{-1}\,\normord{\;\cos(\beta\phi)\;},
  \,\,\,
  Z_1 = \exp\!\left(\frac{\beta^2}{2}\vev{\phi^2}_{0,a}\right),
  \label{eq:normalord}
\end{equation}
where $\vev{\phi^2}_{0,a}$ is the free-field coincident two-point function at finite lattice spacing~\cite{Coleman:1974bu}.

On the infinite chain, with the discretized dispersion relation~\eqref{eq:freedispersion} and the lattice
mass $\mbare$ acting as an infrared regulator, the lattice coincident two-point function is
\begin{align}
  \vev{\phi^2}_{0,a}
  &= \frac{1}{2\pi}\int_{-\pi/a}^{\pi/a}\frac{\d k}{2\omega_k(a)}
  = \frac{\mathrm{K}\!\left(\tfrac{4}{4 + a^2 \mbare^2}\right)}{\pi\sqrt{4 + a^2 \mbare^2}}
  \nonumber\\&= \frac{1}{2\pi}\ln\frac{8}{a \mbare} + \mathcal{O}(a^2 \mbare^2),
  \label{eq:two-point-lat}
\end{align}
with $\mathrm{K}$ the complete elliptic integral of the first kind.

The mass formula in Eq.~\eqref{eq:exactmass} is written in terms of the vertex operator following normalization conventions in conformal field theory (CFT)~\cite{Zamolodchikov:1995xk,Lukyanov:1996jj}
\begin{equation}
  \vev{\normord{\;\;e^{i\beta\phi}\;}\;(x)\,\normord{\;\;e^{-i\beta\phi}\;}\;(0)}_{\mathrm{CFT}}
  = |x|^{-4\alpha}.
  \label{eq:cft-norm}
\end{equation}
The operator is fixed by the coefficient of its two-point function at finite separation $x$, instead
of its value at coincident points. To match the lattice cosine to this normalization, we evaluate the free-field vertex two-point function using Wick's theorem
\begin{align}
  &\vev{e^{i\beta\phi(x)}\,e^{-i\beta\phi(0)}}
  = \nonumber\\
  &\exp\!\Big\{\beta^2\big[\vev{\phi(x)\phi(0)} - \vev{\phi^2}_{0,a}\big]\Big\}.
  \label{eq:wick}
\end{align}
The normalization at separation $x$ thus involves both the coincident function
$\vev{\phi^2}_{0,a}$, which carries the lattice cutoff through~\eqref{eq:two-point-lat}, and the
propagator at finite separation, which carries the cutoff $|x|$. In the continuum, the latter is
the Euclidean massive propagator
\begin{align}
  \vev{\phi(x)\phi(0)}
  &= 
  \int\!\frac{\d^2 k}{(2\pi)^2}\,\frac{e^{i k\cdot x}}{k^2 + \mbare^2}
  = \frac{1}{2\pi}\mathrm{K}_0(\mbare|x|)
  \nonumber\\&= \frac{1}{2\pi}\ln\frac{2e^{-\gamma_E}}{\mbare|x|} + \mathcal{O}(\mbare^2 x^2),
  \label{eq:two-point-cont}
\end{align}
with $\mathrm{K}_n$ the modified Bessel function of the second kind and $\gamma_E$ the
Euler-Mascheroni constant. The mass $\mbare$ plays again the role of an infrared regulator, enters in 
both~\eqref{eq:two-point-lat} and~\eqref{eq:two-point-cont}, but cancels in their difference. 
Inserting both expressions into~\eqref{eq:wick} gives
\begin{equation}
  \vev{e^{i\beta\phi(x)}\,e^{-i\beta\phi(0)}}
  = \left(\frac{a}{\Csch\,|x|}\right)^{\!4\alpha}.
  \label{eq:vertex-twopoint}
\end{equation}
Comparison with the CFT normalization~\eqref{eq:cft-norm} allows us to identify the bare lattice cosine with the conformal one
\begin{equation}
  \cos(\beta\phi)\big|_{\mathrm{lat}}
  = \left(\frac{a}{\Csch}\right)^{\!2\alpha}\normord{\;\cos(\beta\phi)\;}_{\mathrm{CFT}}.
  \label{eq:matching}
\end{equation}
The scheme constant $\Csch$ is determined by the constant pieces of~\eqref{eq:two-point-lat}
and~\eqref{eq:two-point-cont}, and is independent of the regulators $a$ and $|x|$. Writing the
leading logarithms as
\begin{align}
  \vev{\phi^2}_{0,a} &= \frac{1}{2\pi}\ln\frac{c_{\mathrm{lat}}}{a \mbare},
  \nonumber\\
  \vev{\phi(x)\phi(0)} &= \frac{1}{2\pi}\ln\frac{c_{\mathrm{cont}}}{\mbare|x|},
  \label{eq:shortdist}
\end{align}
and using~\eqref{eq:wick}, we obtain
\begin{equation}
    \Csch = \frac{c_{\mathrm{lat}}}{c_{\mathrm{cont}}}
      = \frac{8}{2e^{-\gamma_E}} = 4e^{\gamma_E} \approx 7.12 .
  \label{eq:Cformula}
\end{equation}
Here $c_{\mathrm{lat}} = 8$ comes from the elliptic-integral expansion~\eqref{eq:two-point-lat}
and $c_{\mathrm{cont}} = 2e^{-\gamma_E}$ from the small-argument expansion of the modified
Bessel function~\eqref{eq:two-point-cont}. Since this matching is determined entirely by the free-boson ultraviolet fixed point, 
$\Csch$ is fixed exactly for the lattice regularization and continuum normalization adopted here.
The finite-volume spectrum of the model has been tested extensively against the exact results
with the truncated conformal space approach and the nonlinear integral
equation~\cite{Yurov:1989yu,Feverati:1998va,Feverati:1998dt}. There, the mass-coupling relation is
used as an input to express the CFT-normalized coupling in units of the soliton mass, whereas here
it is tested as an output, since the scheme constant of Eq.~\eqref{eq:Cformula} fixes the
normalization of the lattice cosine independently.

Equations~\eqref{eq:normalord} and~\eqref{eq:matching} fix the renormalized mass parameter
entering~\eqref{eq:exactmass}
\begin{equation}
  \mren^2 = \mbare^2\left(\frac{a}{\Csch}\right)^{\!2\alpha},
  \qquad \Csch = 4e^{\gamma_E}.
  \label{eq:mren}
\end{equation}
Thus, the cutoff dependence resides solely in
the combination $\mbare^2 a^{2\alpha}$, such that the bare mass is tuned as
\begin{equation}
    \mbare(a) = m_{\mathrm{ref}}\left(\frac{a_{\mathrm{ref}}}{a}\right)^{\!\alpha},
    \qquad \alpha = \frac{\beta^2}{8\pi}.
  \label{eq:tuning}
\end{equation}
With the imposed choice $m_{\mathrm{ref}}=a_{\mathrm{ref}}=1$ this fixes
$\mren^2 = \Csch^{-2\alpha}$ independently of $a$. Equivalently, one may keep $\mren=1$
in~\eqref{eq:exactmass} and account for the normalization through the constant factor
\begin{equation}
  \Mzam = \Msol\big|_{\mren=1}\, \mathcal{N}(\beta),
  \qquad
  \mathcal{N}(\beta) = \Csch^{-\xi} = \bigl(4e^{\gamma_E}\bigr)^{-\xi},
  \label{eq:schemecorr}
\end{equation}
which follows from $\Msol \propto \mren^{1/(1-\alpha)}$ together with $\xi = \alpha/(1-\alpha)$.

The results presented below follow a three-step matching prescription. We first set $\mbare(a)$
from~\eqref{eq:tuning} with $m_{\mathrm{ref}}=1$ at $a_{\mathrm{ref}}=1$. We then compute the
two vacua and the QP soliton mass $\Mqp(a)$. Finally, we compare with $\Mzam$
from~\eqref{eq:exactmass} evaluated at $\mren$ of~\eqref{eq:mren}, or, equivalently, by applying the
constant factor in~\eqref{eq:schemecorr}. Since both the scheme constant and Zamolodchikov's
formula are exact, the ratio $\Mqp/\Mzam$ approaches unity as $a \to 0$. The
deviation at finite $a$ is controlled by the lattice discretization together with the bond-dimension and Fock-space truncations, all of which vanish in the appropriate limits.

\section{Numerical Results}
\label{sec:results}

We first test the static spectrum and its continuum limit, then turn to real-time soliton
antisoliton scattering. Unless stated otherwise, the spectrum scans use
$\beta\in\{1.0,1.5,2.0,2.5,3.0\}$, Fock cutoff $d=90$, bond dimension $\chi=64$, and the
scheme-corrected mass~\eqref{eq:schemecorr}. The spectrum is obtained on the infinite chain,
such that the thermodynamic limit is exact there and no finite-volume error enters. Only the
scattering simulations place the state on a finite lattice, with the boundary conditions
specified in \appref{app:tn}. The tensor network constructions are
described in \appref{app:tn}.

\subsection{Solitons, Breathers and the Continuum Limit}
\label{sec:results:spectrum}

\begin{figure}[t]
\centering
\includegraphics[width=\columnwidth]{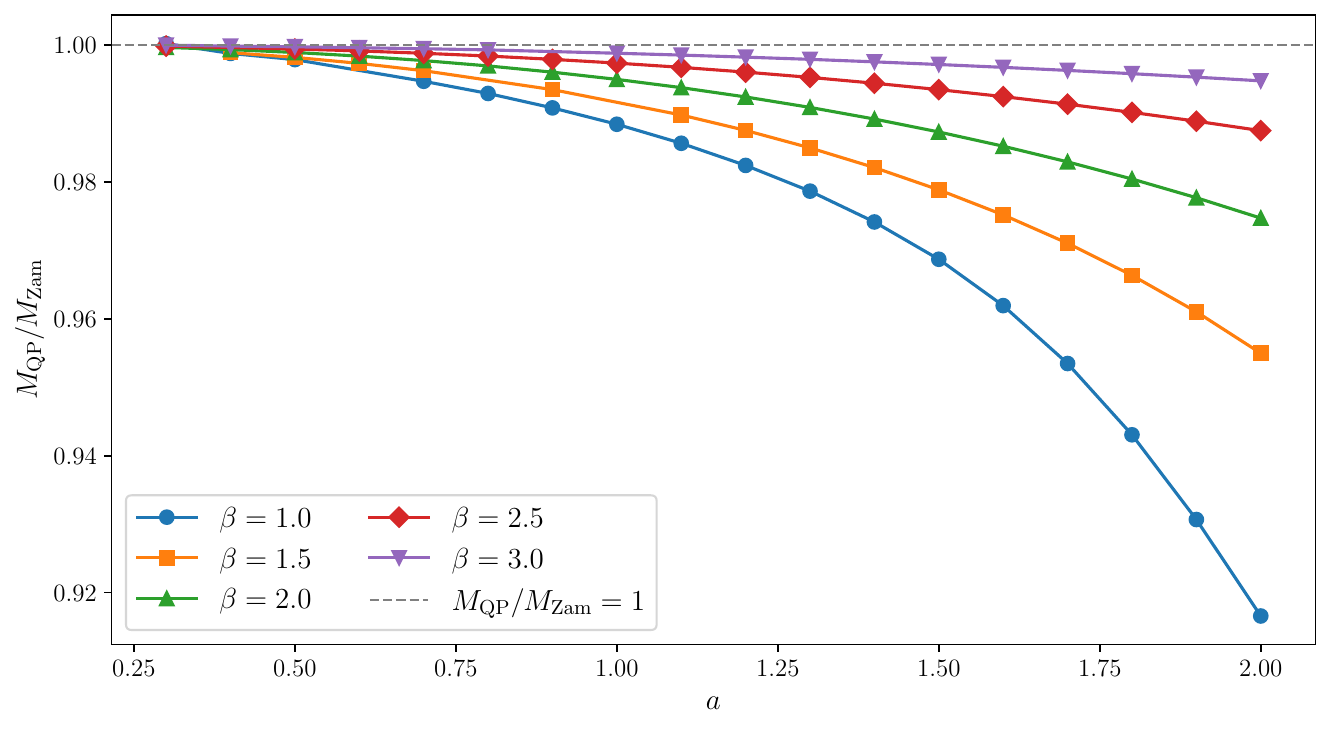}
  \caption{Ratio of the quasiparticle soliton mass to the scheme-corrected Zamolodchikov
           result~\eqref{eq:schemecorr} as a function of the lattice spacing, for $d = 90$ and
           $\chi = 64$.}
  \label{fig:solitonmass}
\end{figure}

\figref{fig:solitonmass} shows the ratio $\Mqp/\Mzam$ as a function of the lattice spacing for five
couplings $\beta$. The ratio approaches unity as $a \to 0$ in every case, which is the parameter-free
statement of \secref{sec:renorm}. We further note that the more modest value for the Fock cut-off of $d=25$ used in the scattering simulation below is sufficient for most of the data points in the above figure. Only for $\beta\leq 1.5$ a higher Fock cut-off is required. Convergence is from below and monotonic for
$\beta \geq 1$, with the residual deviation at $a = 2$ growing as the coupling decreases.
At small $\beta$, the two vacua are separated by $2\pi/\beta$, which is a large separation in field space in units of the oscillator length, so the same Fock cutoff resolves the soliton progressively worse. Large couplings show the opposite behavior, converging correctly but slowly in $a$.

\begin{figure*}[t]
\includegraphics[width=.85\textwidth]{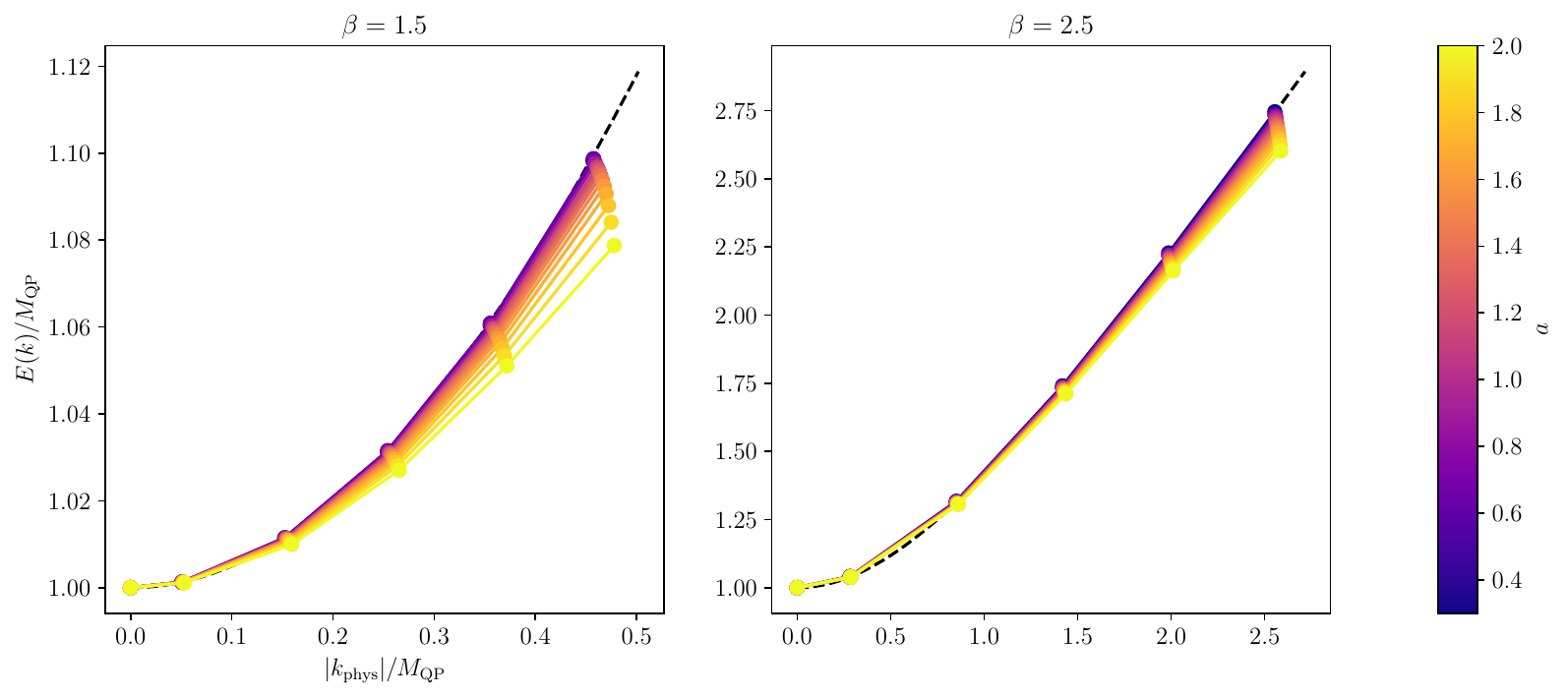}
\caption{Normalized soliton dispersion relation $E(k)/\Mqp$ at two representative couplings $\beta=1.5$ and $\beta=\sqrt{2\pi}$ versus physical momentum for all scanned lattice spacings. The dashed curve corresponds to the continuum result
         $\sqrt{1+(|k_{\rm phys}|/\Mqp)^2}$.}
\label{fig:dispersion_grid}
\end{figure*}

Next, \figref{fig:dispersion_grid} shows the normalized soliton dispersion relation $E(k)/\Mqp$ for two reference values in $\beta$, together with the relativistic continuum reference $\sqrt{M^2+k^2}$ at the extracted soliton mass. In the limit $a\to0$ we observe restoration of Lorentz invariance and we recover the correct continuum limit.

Breather masses are computed with the same-vacuum QP ansatz described in \appref{app:tn} and in \tabref{tab:breather_masses} we compare them with
Eq.~\eqref{eq:breathermasses} evaluated at the scheme-corrected soliton mass. At $a = 0.3$, $d=90$ and $\chi = 64$, the lightest breather
reproduces the exact tower to better than $0.2\%$ for all values of $\beta$, except for $\beta=1.0$, where the third breather deviates by about 25\%. The heavier states sit closer to the two-soliton threshold $2\Msol$ and are correspondingly more sensitive to the Fock cutoff. For larger couplings, the interaction becomes repulsive and the spectrum does not contain any breathers.

\begin{table}[htbp]
  \centering
  \begingroup
  \small
  \renewcommand{\arraystretch}{1.15}
  \begin{tabular*}{\columnwidth}{@{\extracolsep{\fill}}ccccc@{}}
    \toprule
        $\beta$ & $n$ & $M_B$ & $M_{\text{Br}}$ & Ratio \\
        \midrule
        1.0 & 1 &   0.917     &   0.916     &   1.000     \\
            & 2 &   1.844     &   1.829     &   1.009     \\
            & 3 &   2.053     &   2.733     &   0.751     \\
        \midrule
        1.5 & 1 &   0.810     &   0.810     &   1.001     \\
            & 2 &   1.605     &   1.599     &   1.004     \\
            & 3 &   1.779     &   2.351     &   0.757     \\
        \midrule
        2.0 & 1 &   0.658     &   0.657     &   1.000     \\
            & 2 &   1.259     &   1.257     &   1.002     \\
            & 3 &   1.488     &   1.747     &   0.852     \\
    \midrule
        2.5 & 1 &   0.467     &   0.467     &   1.000     \\
            & 2 &   0.812     &   0.810     &   1.001     \\
            & 3 &   0.980     &   0.940     &   1.043     \\
        \midrule
        3.0 & 1 &   0.259     &   0.259     &   1.001     \\
        \bottomrule
  \end{tabular*}
  \endgroup
  \caption{Comparison of numerically calculated breather masses $M_B$ at lattice spacing $a = 0.3$ together with their theoretical values $M_{\text{Br}} = 2 M_{\text{sol}} \sin(n \pi \xi / 2)$ across various coupling parameters $\beta$. The values are rounded to three significant digits.}
  \label{tab:breather_masses}
\end{table}

\subsection{Soliton antisoliton Scattering}
\label{sec:results:scattering}

The scattering runs are performed at the reflectionless point $\beta = \sqrt{2\pi}\approx 2.5$,
with $m_{\rm ref} = 1$ at $(d, \chi) = (25, 64)$ for $a = 1.00$ and $(30, 80)$ for $a = 0.75$, the
largest bond dimension available at each spacing. 
Gaussian wave packets of physical envelope width $\sigma_x = 8$ are prepared for the soliton and
the antisoliton with equal and opposite momenta, placed symmetrically about the chain center at
an initial separation of $5.0\,\sigma_x$, and evolved with two-site TDVP. All physical quantities are
held fixed across lattice spacings, such that runs at different $a$ describe the same physical kinematics.
\tabref{tab:runs} lists the parameters of four runs.

\begin{table}[htbp]
  \centering
  \begingroup
  \small
  \renewcommand{\arraystretch}{1.15}
  \begin{tabular*}{\columnwidth}
    {@{\extracolsep{\fill}}ccccccc@{}}
    \toprule
    Run & $a$ & $p$ & $N$ & $\sigma_x$ & $\chi$ & $t_{\mathrm{coll}}$ \\
    \midrule
    1 & 1.00 & 0.80 & 120 & 8.00 & 64 & 23.2 \\
    2 & 0.75 & 0.80 & 160 & 8.00 & 80 & 23.5 \\
    3 & 1.00 & 0.60 & 120 & 8.00 & 64 & 25.4 \\
    4 & 0.75 & 0.60 & 160 & 8.00 & 80 & 25.7 \\
    \bottomrule
  \end{tabular*}
  \endgroup
  \caption{Scattering runs. Each is evolved to $t=60$ and paired with an
    independent single-soliton run at identical
    $(\beta,p,m,a,\chi,d)$ that supplies the free reference trajectory.}
  \label{tab:runs}
\end{table}

\begin{figure}[t]
    \centering
    \includegraphics[width=\columnwidth]{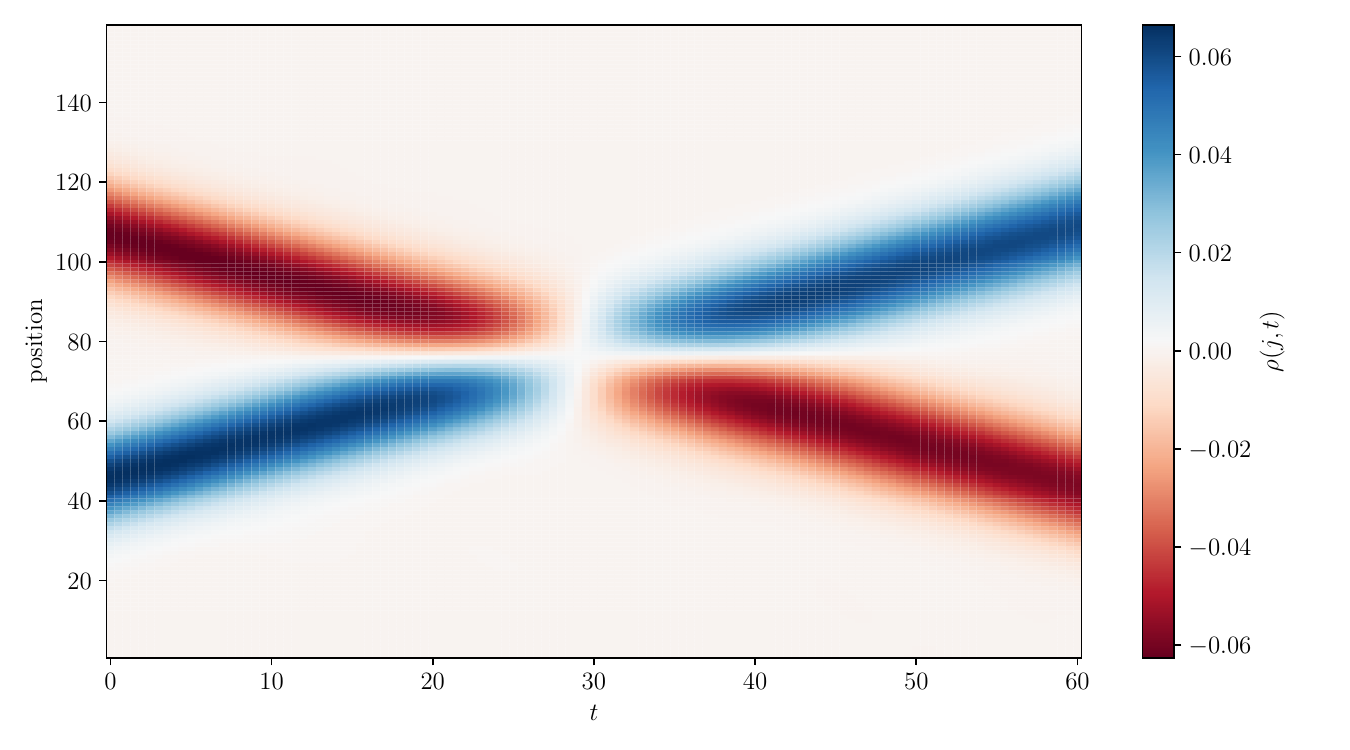}
    \caption{Topological charge density $\rho(j,t)$ for run 4 of \tabref{tab:runs} at
             $a = 0.75$, $p = 0.60$, $N = 160$. The soliton carries positive topological charge density $\rho > 0$ while the antisoliton
             carries $\rho < 0$. The trajectories confirm the expected transmissive behavior.}
    \label{fig:charge_density}
\end{figure}

To discriminate between soliton and antisoliton at every time step, we measure the topological charge
density $\rho(j,t)$ instead of the energy density. The advantages of this choice are twofold.
First, the charge density is sign-definite and allows for a clear discrimination between soliton and
antisoliton. Secondly, the energy density contains an infinite polynomial of the field due to
the cosine term in Eq.~\eqref{eq:latticeH}, making it much more susceptible to noise.
\figref{fig:charge_density} shows the time evolution of the charge density for run 4.
The positive and negative charge profiles approach, merge near $t_{\mathrm{coll}}$, and emerge with their
signs and velocities intact after the collision, but a finite displacement relative to free propagation.

The positions of the soliton and antisoliton are extracted as the charge-weighted centroid of the corresponding charge profile above a threshold
set at a fraction $\theta_\rho = 0.25$ of its peak
\begin{align}
    x_{S/\bar{S}}(t)
    &= \frac{\displaystyle\sum_{j\in J^\pm} \bigl(j+\tfrac{1}{2}\bigr)
            \bigl[\rho^{\pm}(j,t) - \theta_\rho\,\rho^{\pm}_{\mathrm{peak}}(t)\bigr]}
           {\displaystyle\sum_{j\in J^\pm}
            \bigl[\rho^{\pm}(j,t) - \theta_\rho\,\rho^{\pm}_{\mathrm{peak}}(t)\bigr]},
    \nonumber\\
    \rho^{\pm} &= \max(\pm\rho,\,0),
    \label{eq:centroid}
\end{align}
where $\rho^{\pm}_{\mathrm{peak}}(t)=\max_j\rho^{\pm}(j,t)$ is the peak height of the corresponding charge profile at time $t$,
$J^\pm$ is the contiguous set of sites around the peak on which
$\rho^\pm > \theta_\rho\rho^\pm_{\mathrm{peak}}$, and the shift $j \to j+\tfrac{1}{2}$ accounts for
$\rho$ being a two-site operator defined on the link $(j, j+1)$.

\begin{figure*}[t]
    \centering
    \includegraphics[width=.85\textwidth]{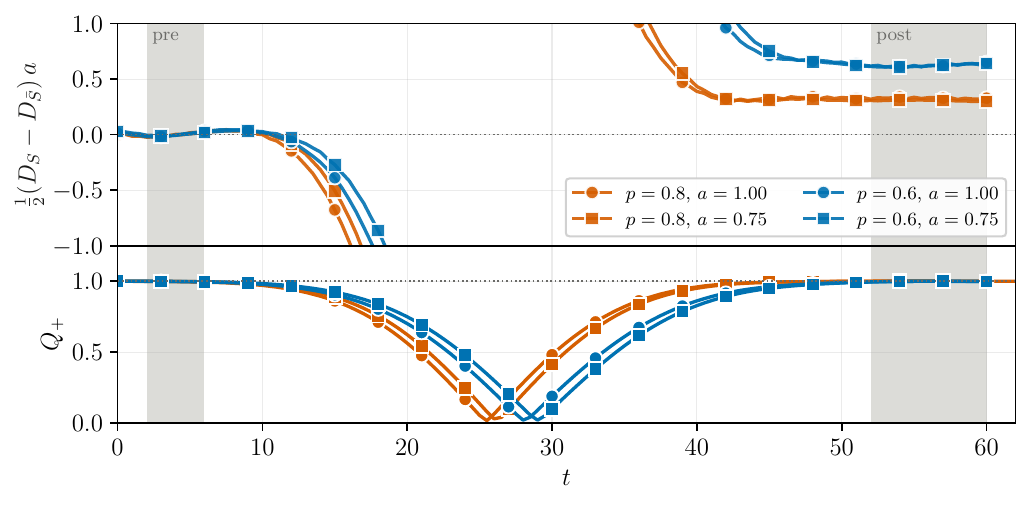}
    \caption{Antisymmetric displacement $\tfrac{1}{2}(D_S - D_{\bar S})a$ relative to the free
             reference, referenced to the pre-collision window, and the positive charge $\Qpos$. Shaded bands mark the pre-collision window $[2,6]$ and the
             post-collision window $[52,60]$.}
    \label{fig:tracks}
\end{figure*}

The asymptotic window is chosen from the positive charge,
$\Qpos(t) = a\sum_j\max(\rho,0)$, which equals unity for an isolated soliton. Requiring
$|\Qpos(t) - 1| < 5\times10^{-3}$ for the remainder of the evolution is satisfied from
$t \simeq 46$ in runs 1 and 2 and from $t = 52$ in runs 3 and 4, such that the post-collision
window $[52, 60]$ describes asymptotic states for all four runs. The pre-collision window is taken as $[2, 6]$, allowing TDVP to relax the wave packet further, but while the pair is still $4.2\,\sigma_x$ apart. In \figref{fig:tracks} we illustrate the choice of window through the soliton tracks.

The scattering displacement is measured against a free reference trajectory taken from an
independent single-soliton run at identical parameters. Writing
\begin{equation}
    D_{S/\bar{S}}(t) = x_{S/\bar{S}}^{\mathrm{scatt}}(t)
                     - x_{S/\bar{S}}^{\mathrm{free}}(t),
\end{equation}
the collision-induced shift is the change in $D$ between the two windows
\begin{equation}
    \Delta x_{S/\bar{S}}
    = \bigl\langle D_{S/\bar{S}}\bigr\rangle_{\mathrm{post}}
    - \bigl\langle D_{S/\bar{S}}\bigr\rangle_{\mathrm{pre}}.
\end{equation}
Subtracting the pre-collision mean removes a static offset between the two runs, which
originates in the residual gauge freedom of the wave packets construction described in
\appref{app:tn}, and it cancels the common wave packets spreading and truncation drift. The
physical displacement is then given by the antisymmetric combination
\begin{equation}
\label{eq:displacement_phys}
    \dxphys
    = \tfrac{1}{2}\bigl(\Delta x_S - \Delta x_{\bar{S}}\bigr),
\end{equation}
which reduces to $\Delta x_S = -\Delta x_{\bar{S}}$ for a perfectly symmetric collision, while
the symmetric combination
\begin{equation}
\label{eq:displacement_cm}
    \dxcm
    = \tfrac{1}{2}\bigl(\Delta x_S + \Delta x_{\bar{S}}\bigr)
\end{equation}
carries no physical displacement and serves as a diagnostic for residual systematics.

Equation~\eqref{eq:displacement_analytical} predicts $\dxphys$ for momentum eigenstates. The
simulated states are Gaussian wave packets of width $\sigma_x = 8$, corresponding to a momentum
spread $\sigma_k = 0.088$. Since $\Delta x(\theta_{12})$ is a convex function over that range, the
wave packet average
\begin{align}
  \langle\Delta x\rangle
  &= \int\!\d k_1\!\int\!\d k_2\; w(k_1)\,w(k_2)\,\Delta x(k_1,k_2),
  \nonumber\\ w &\propto |\tilde f|^2,
  \label{eq:dx_wp}
\end{align}
does not coincide with the eigenstate value but instead corresponds to a roughly 4\% correction. In
\tabref{tab:dx} we quote both values for completeness, though the result of our simulation should be compared to the latter.
\begin{table}[htbp]
  \centering
  \begingroup
  \small
  \renewcommand{\arraystretch}{1.2}
  \begin{tabular*}{\columnwidth}
    {@{\extracolsep{\fill}}cccccc@{}}
    \toprule
    Run
      & $\theta_{12}$
      & $\dxphys$
      & $\Delta x_{\mathrm{eig}}$
      & $\langle\Delta x\rangle$
      & Ratio \\
    \midrule
    1 & 2.609 & 0.329(12) & 0.277 & 0.286 & 1.15(4) \\
    2 & 2.607 & 0.307(9)  & 0.278 & 0.287 & 1.07(3) \\
    3 & 2.133 & 0.625(16) & 0.544 & 0.566 & 1.10(3) \\
    4 & 2.131 & 0.622(12) & 0.545 & 0.567 & 1.10(2) \\
    \bottomrule
  \end{tabular*}
  \endgroup
  \caption{Measured scattering displacement compared with the exact
    prediction, in physical length units. Here,
    $\Delta x_{\mathrm{eig}}$ is Eq.~\eqref{eq:displacement_analytical}
    for momentum eigenstates, $\langle\Delta x\rangle$ is its wave packet
    average in Eq.~\eqref{eq:dx_wp}, and the ratio is
    $\dxphys/\langle\Delta x\rangle$. The relative rapidity
    $\theta_{12}=2\arcsinh(p/\Mqp)$ differs between lattice spacings at
    fixed $p$ only through $\Mqp$.
    Runs at the same momentum therefore nearly coincide on the rapidity
    axis of \figref{fig:displacement}.}
  \label{tab:dx}
\end{table}
\begin{figure*}[ht]
    \centering
    \includegraphics[width=.9\textwidth]{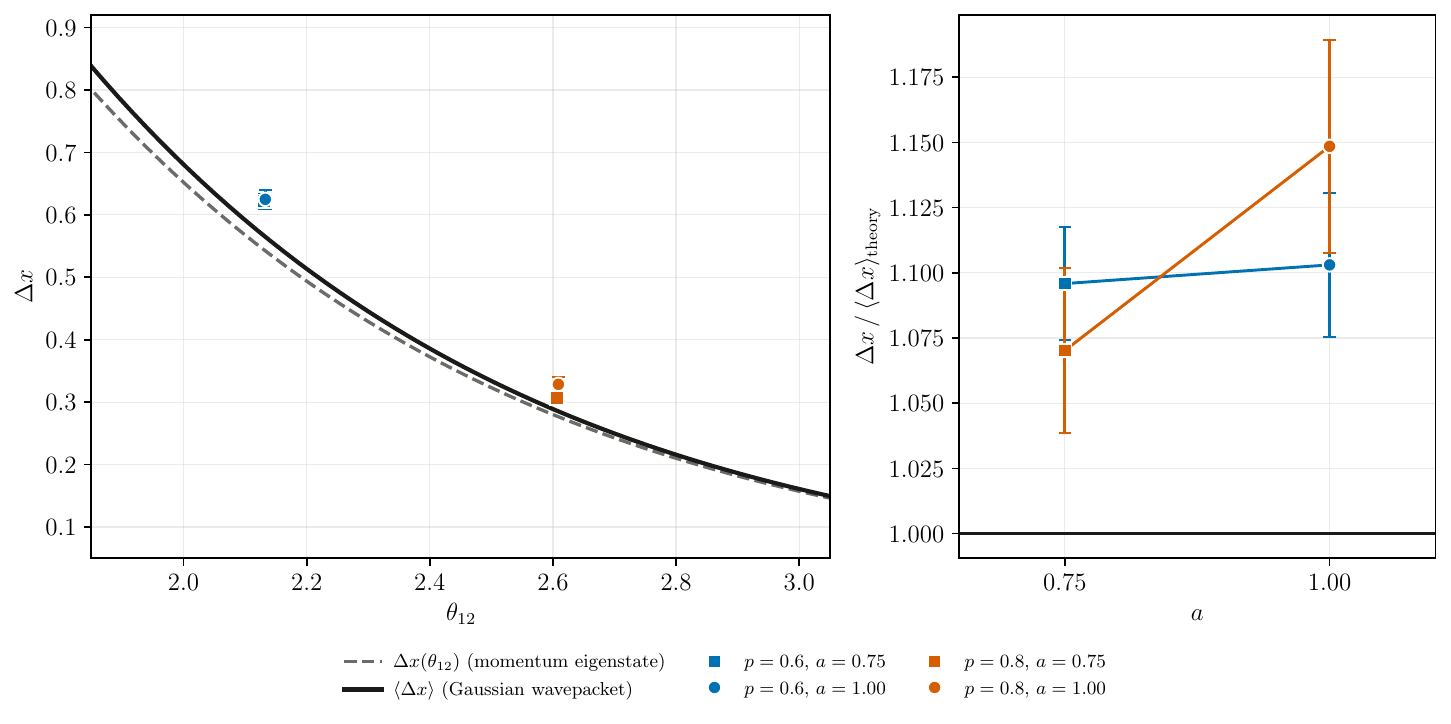}
    \caption{Measured displacement versus relative rapidity, compared with
             Eq.~\eqref{eq:displacement_analytical} for momentum eigenstates (dashed) and with
             its packet average~\eqref{eq:dx_wp} at $\sigma_k = 0.088$ (solid). The right panel
             shows the ratio to the packet average as a function of lattice spacing. Color encodes the
             momentum and the marker the lattice spacing.}
    \label{fig:displacement}
\end{figure*}
The quoted error is obtained in quadrature from three contributions. First, our analysis shows a quite strong sensitivity to the choice of pre-collison window as determined in \figref{fig:tracks}. We thus choose six different slicings of that window and the corresponding error $e_{\rm pre}$ corresponds to half of the measured spread. Secondly, we vary the threshold $\theta_\rho$ in \eqref{eq:centroid} between $10-50\%$ and denote $e_{\rm thr}$ as half the measured spread for this variation. Lastly, the third contribution is given by $e_{\rm scat}$ and corresponds to fit residuals of the tracks. To summarize, we compute the error as
\begin{equation}
    {\rm err}=\sqrt{e_{\rm pre}^2+e_{\rm thr}^2+e_{\rm scat}^2}
\end{equation}
Out of the three, the pre-window term gives the largest contribution. This could possibly be improved by choosing a larger system and correspondingly further separated initial state wave packets.

In \figref{fig:displacement} we show the numerically extracted displacements together with the eigenstate continuum prediction, as well as the momentum smeared value in \eqref{eq:dx_wp}. As expected, we see an improvement as we get closer to the continuum limit $a\to0$ and by increasing the bond dimension. The agreement with the theory prediction is better than 10\% already at the quite modestly chosen bond dimension $\chi=80$ and Fock cut-off of $d=30$ for $a=0.75$, which we have chosen as a trade-off between computational resources and accuracy. Based on the results of the continuum value for the soliton mass and its dispersion relation, the excess is most likely due to finite bond dimension effects at the chosen lattice spacings. Additionally, the initial state corresponds to a product of two wave packets, rather than individual asymptotic eigenstates, which introduces an unphysical offset already at the initialization stage. Finally, the gauge offset of the wave packet construction described in \appref{app:tn} leads to a small excess energy that propagates through the lattice and distorts the physical propagation of the scattering configuration. The diagnostic $\dxcm$ described in \eqref{eq:displacement_cm} vanishes for an ideal collision but is found to be around 5-10\%.

\section{Conclusions and Outlook}
\label{sec:conclusions}

We have used bosonic tensor networks to connect the lattice
sine-Gordon Hamiltonian quantitatively to its continuum
theory. Matching the lattice vertex operator to its
conformal normalization at the free-boson ultraviolet
fixed point determines the renormalization independently
of the numerical spectrum. This establishes a
parameter-free comparison with Zamolodchikov's exact
mass-coupling relation, which the soliton mass approaches
in the continuum limit throughout the studied parameter
range. The recovery of the relativistic soliton dispersion
and the two lightest breather masses at the sub-percent level
provides complementary tests of the bosonic formulation
and its continuum matching.

Real-time collisions of Gaussian soliton-antisoliton
wave packets near a reflectionless point reproduce the
expected transmission behavior. Tracking the topological
charge density allows us to extract the Wigner spatial displacement,
which we compare with the exact continuum prediction
from the momentum derivative of the transmission phase,
averaged over the packet momentum distributions.
The measured displacements follow the predicted rapidity
dependence, with deviations from the continuum values
by less than $10\%$. The lattice-spacing dependence suggests
improved agreement toward the continuum, although the
uncertainties remain sizable. Larger-scale simulations
with increased local Fock cutoffs would extend this
comparison to finer lattices and sharpen the continuum
test, albeit at significantly greater computational cost.

Our bosonic simulations provide a starting point for
studying reflection and transmission away from the
reflectionless point, collisions involving breathers,
and nonintegrable deformations for which exact scattering
predictions are generally unavailable. Beyond sine-Gordon
theory, the renormalization analysis illustrates how
the regulator-dependent normalization of lattice
operators can be established independently of the
calculation of physical observables. This separation
offers useful guidance for connecting other scalar
Hamiltonian lattice theories to their continuum limits
and can inform related renormalization studies in
lattice gauge theories. The matching conditions and
required counterterms are theory dependent, but the
strategy of independently fixing the bare parameters
and testing multiple physical observables is broadly
applicable.

Finally, the explicit oscillator representation connects
directly to continuous-variable quantum simulation,
where local bosonic modes encode the field and its
conjugate momentum. The spectrum, dispersion, and
scattering observables studied here provide classical
benchmarks for hybrid qubit-qumode implementations
of the same lattice Hamiltonian~\cite{Rainaldi:2025ymn}.
Together with the continuum matching, these benchmarks
allow state preparation and real-time evolution to be
assessed in terms of physical particle properties
and interactions.


\section*{Data and Code Availability}

To promote open and reproducible research, the data and computer code used to generate the numerical results and figures in this paper are publicly available via Zenodo \cite{zenodo22812679} and GitHub \cite{sine-gordon-mpskit}.


\begin{acknowledgments}
We would like to thank Dima Kharzeev and Jake Montgomery for helpful discussions. F.H.\ is funded by the Austrian Science Fund (FWF) [10.55776/J4854]. TR and FR are supported by the DOE, Office of Science, Office of Nuclear Physics, Early Career Program under contract No. DE-SC0025881. The computational results presented have been achieved in part using the Vienna Scientific Cluster (VSC). This research used resources of the National Energy Research Scientific Computing Center (NERSC), a DOE Office of Science User Facility using NERSC award NP-ERCAP0037787. The authors would like to thank Stony Brook Research Computing and Cyberinfrastructure, and the Institute for Advanced Computational Science at Stony Brook University, for access to the SeaWulf computing system, made possible by grants
from the National Science Foundation (\#1531492 and Major Research Instrumentation award \#2215987), with
matching funds from Empire State Development’s Division of Science, Technology and Innovation (NYSTAR)
program (contract C210148).
\end{acknowledgments}

\appendix

\section{Uniform Bosonic Tensor Networks}
\label{app:tn}

This appendix collects the tensor network constructions used in the main text. The framework is
adapted from Refs.~\cite{Zauner-Stauber:2017eqw,Vanderstraeten:2019voi,Milsted:2020jmf,VanDamme_2021},
to which we refer the interested reader for further details.
All computations were performed using the \texttt{MPSKit} library~\cite{Devos_MPSKit_2026} for \texttt{Julia}~\cite{bezanson2017julia}.

The two degenerate vacua are obtained as uniform MPS using the VUMPS algorithm~\cite{Zauner-Stauber:2017eqw}.
A spatially localized excitation on top of them is built from the vacuum tensors $A_L$ and
$A_R$, which are held fixed, together with an excitation tensor $B$ placed at site $j$
\begin{align}
    &\ket{\chi_j(A_L,B,A_R)}
    = \nonumber\\
    &\sum_{\vec{s}} v_L^\dagger
      \left(\prod_{i=-\infty}^{j-1}A_L^{s_i}\right)
      B^{s_j}
      \left(\prod_{i=j+1}^{\infty}A_R^{s_i}\right)
      v_R\ket{\vec{s}},
  \label{eq:qpansatz}
\end{align}
where $\ket{\vec{s}}$ runs over the physical basis of the chain and $v_{L/R}$ fixes the norm and
boundary conditions. Momentum eigenstates are given by Fourier modes of these localized states
\begin{equation}
  \ket{\chi(A_L,B,A_R,p)} = \sum_j e^{i p j}\,\ket{\chi_j(A_L,B,A_R)}.
  \label{eq:qpmomentum}
\end{equation}
The energy $E(p)$ and the optimal exciation tensor $B$ follow from projecting the Hamiltonian onto this ansatz
space and solving the resulting eigenvalue problem. Since the ansatz is variational in the
tangent space of the vacuum, the resulting masses are upper bounds at fixed bond dimension $\chi$.

The topological charge is fixed by the choice of vacua in Eq.~\eqref{eq:qpansatz}. 
A soliton is a domain wall between two distinct vacua, obtained by
taking $A_L = A_L^{(0)}$, the left-canonical tensors of the vacuum with $\vev{\phi}=0$, and
$A_R = A_R^{(1)}$, the right-canonical tensors of the vacuum with $\vev{\phi}=2\pi/\beta$. This
gives $Q=+1$, and the resulting state is denoted as $\ket{\kappa_j}$. Then antisoliton $\ket{\bar\kappa_j}$ with $Q=-1$
is obtained by exchanging the two vacua. The breather modes $\ket{\kappa\bar\kappa_j}$ have vanishing topological charge,
and therefore belong to the same topological sector as the trivial vacuum. Their quasiparticle ansatz
is thus constructed by using the same vacuum on both sides $A_L = A_R$. Soliton, antisoliton, and breather thus share the
same variational form and differ only in the pair of vacua used in their construction. The two vacua are the
degenerate vacuum solutions obtained via VUMPS related by the shift $\phi\to\phi+2\pi/\beta$.

A wave packet localized at $x_0$ with mean momentum $p_0$ and momentum width $\sigma_p$
is obtained by the superposition
\begin{equation}
  \ket{\Psi_\text{wp}(p_0, \sigma_p, x_0)}
  \propto \sum_p
    f(p)
    e^{-i p x_0}
    \ket{\chi(p)},
  \label{eq:wavepackets}
\end{equation}
where $\ket{\chi(p)}$ is any of $\ket{\kappa(p)}$, $\ket{\bar{\kappa}(p)}$ or
$\ket{\kappa\bar{\kappa}(p)}$ and
\begin{equation}
    f(p)=e^{-(p-p_0)^2/(2\sigma_p^2)}
\end{equation}
is the associated Gaussian weight. Since the Fourier transform of a Gaussian is again a
Gaussian, the weights in Eq.~\eqref{eq:wavepackets} can equivalently be applied in position
space. For the numerical simulation, we neglect the variation of $B(p)$ over the momentum
support of the packet and use a single excitation tensor $B(p_0)$, which gives
\begin{align}
  \ket{\Psi_\text{wp}} \propto
  \sum_{j=1}^{N} f_j\ket{\chi_j(B(p_0))},\nonumber\\
  f_j=\mathcal{N}e^{-(j-x_0)^2/(2\sigma_x^2)}e^{ip_0j},
  \qquad \sigma_x=\sigma_p^{-1},
  \label{eq:wave packet-position}
\end{align}
up to an overall phase. On the finite chain, the coefficients are evaluated and normalized
over the chain length $j=1,\ldots,N$. The resulting state is then compressed to bond dimension $\chi$ by singular
value decomposition.

Such a superposition is again a single MPS. For one soliton, the site tensors take the
$2\times2$ block form described in Ref.~\cite{Milsted:2020jmf}
\begin{equation}
  A_j^{s} =
  \begin{pmatrix}
    A_L^{s} & f_j B^{s}\\
    0       & A_R^{s}
  \end{pmatrix},
  \label{eq:block2}
\end{equation}
of bond dimension $2D$ when $A_L$, $A_R$ and $B$ each carry bond dimension $D$. A
soliton-antisoliton pair needs a central vacuum $A_C$ between the two excitations, and the corresponding tensor is given by the $3\times3$ block
\begin{equation}
  A_j^{s} =
  \begin{pmatrix}
    A_L^{s} & f_j B_\kappa^{s} & 0 \\
    0       & A_C^{s}     & g_j B_{\bar\kappa}^{s} \\
    0       & 0           & A_R^{s}
  \end{pmatrix},
  \label{eq:block3}
\end{equation}
where
\begin{align}
  f_j=&\,\mathcal{N}_\kappa e^{-(j-x_\kappa)^2/(2\sigma_x^2)}e^{ip_0j},\nonumber\\
  \qquad
  g_j=&\,\mathcal{N}_{\bar\kappa}e^{-(j-x_{\bar\kappa})^2/(2\sigma_x^2)}e^{-ip_0j}.
  \label{eq:two-wave packet-weights}
\end{align}
The upper-triangular structure ensures a transition from $A_L$ to $A_C$ proceeds through
$B_\kappa$ and a subsequent transition from $A_C$ to $A_R$ through $B_{\bar\kappa}$.
To obtain a symmetric scattering configuration, we use equal spatial widths, opposite mean momenta, and centers placed symmetrically about the midpoint of the chain. 

Note that the Fourier mode~\eqref{eq:qpmomentum} is invariant under
$B^{s} \to B^{s} + A_L^{s} X - e^{-ip} X A_R^{s}$ for any matrix $X$, but the localized
states~\eqref{eq:qpansatz}, and hence the wave packets built from them, are not. We fix $X$ using the reflection-symmetric condition of Ref.~\cite{VanDamme_2021} as adapted in
Ref.~\cite{Milsted:2020jmf}, which minimizes the overlap of $B$ with both $A_L$ and $A_R$. 
As discussed in Ref.~\cite{Milsted:2020jmf}, this induces a finite offset from the initial position of the wave packet, which we remove by referencing the displacement to the pre-collision window, as discussed in \secref{sec:results:scattering}.

An additional complication arises when a configuration of non-zero topological charge, as is
the case for the freely propagating reference runs described in the main text, is placed on the finite chain.
Since the solitons interpolate between distinct vacua, the endpoints of the
finite chain in Eq.~\eqref{eq:latticeH} must be treated explicitly. We impose Dirichlet
conditions through ghost sites beyond each end of the chain, pinned at the vacuum value adjacent
to that end. In particular, for the single-soliton reference run the adjacent vacua are given by $0$ and $2\pi/\beta$, respectively. 
With the ghost field matched to the neighboring vacuum, the time evolution using the modified 
Hamiltonian does not drive the state into the trivial vacuum, while leaving the underlying physics intact.

\bibliographystyle{unsrt}
\bibliography{bib/references}

\end{document}